\documentclass[12pt]{article}
\usepackage{epsfig}     
\def\singlespace{\baselineskip=14.4pt}
\def\doublespace{\baselineskip=20pt}

\begin{document}


\title{\vspace*{-3cm}\rightline{\normalsize Preprint Saclay-T99/095}\vspace*{2cm}
       Spin Glass Overlap Barriers in Three and Four
       Dimensions\thanks{This research was partially funded by the
                  Department of Energy under contracts DE-FG02-97ER41022
                  and DE-FG05-85ER2500.}
      }
%

\author{Bernd A. Berg\\
\small Department of Physics, The Florida State University,
       Tallahassee, FL~32306, USA.\\
\small and\\
\small Supercomputer Computations Research Institute,\\
\small The Florida State University, Tallahassee, FL~32306, USA.\\
\small E-mail: \texttt{berg@scri.fsu.edu}\\[1.5ex]
Alain Billoire\\
\small CEA/Saclay, Service de Physique Th\'eorique,
       91191 Gif-sur-Yvette, France\\
\small E-mail: \texttt{billoir@spht.saclay.cea.fr}\\[.5ex]
              and \\[.5ex]
Wolfhard Janke\\
\small Institut f\"ur Theoretische Physik, Universit\"at Leipzig,
       04109 Leipzig, Germany\\
\small E-mail: \texttt{wolfhard.janke@itp.uni-leipzig.de}
}
\date{Revised: January 8, 2000}
\maketitle

\begin{abstract}
For the Edwards-Anderson Ising spin-glass model in three and four
dimensions ($3d$ and $4d$) we have performed high statistics Monte Carlo
calculations of those free-energy barriers $F^q_B$ which are visible in
the probability density $P_{\cal J}(q)$ of
the Parisi overlap parameter $q$. The calculations rely on the recently
introduced multi-overlap algorithm. In both dimensions, within the
limits of lattice sizes investigated, these barriers are found to be
non-self-averaging and the same is true for the autocorrelation times
of our algorithm. Further, we present evidence that barriers hidden in
$q$ dominate the canonical autocorrelation times.
\end{abstract}

\vfill
\newpage   

\doublespace


\section{Introduction}

Spin glasses (for reviews see references \cite{Young97,Fi91,Me86,Bi86})
constitute an important class of materials whose
low-temperature state is a frozen disordered one. In order to produce
such a state, there must be randomness and frustration among the
different interactions between the spins (magnetic moments).
Frustration means that no single spin configuration is favored by
all interactions. In real materials such competing interactions are
for instance created by magnetic impurity moments.
The study of spin glasses developed essentially since the middle of
the 1970's and is based on three approaches: experiment, theory and
computer simulation.

Experimentally it is not hard to find spin glasses~\cite{Fi91}. One 
kind of widely studied systems consists of dilute solutions of 
transition
metal magnetic impurities in noble hosts. The impurity moments
produce  a magnetic polarization of the host metal conduction electrons
which is positive at some distances and negative at others.
Because of the random placements of the impurities they have random,
competing interactions with one another. Spin glass states have also
been found in magnetic insulators and amorphous alloys. Properties
analogous to those of spin glasses, with the electric dipole moment
playing the role of the magnetic one, have been seen in 
ferroelectric-antiferroelectric mixtures. The universal behavior of
the observed phenomena is a major reason for the interest in these
systems.

 A freezing temperature $T_c$ may be 
defined by a cusp in the ac susceptibility and has, for instance, been 
studied for Cu-0.9\%~Mn~\cite{Mu81}. Below this transition temperature
characteristic non-equilibrium phenomena are observed. A typical
experiment is the measurement of the remanent magnetization, 
see~\cite{Gr87} for a study of $({\rm Fe}_{0.15} {\rm Ni}_{0.85})_{75} 
{\rm P}_{16} {\rm B}_6 {\rm Al}_3$. 
A spin-glass sample is rapidly cooled in a magnetic field
to a temperature below the transition temperature and the observation 
is that the decay of the magnetization depends on the waiting time 
after which the field is switched off. This phenomenon 
is called aging and has also been found in
other disordered or amorphous systems such as structural glasses,
polymers, high-temperature superconductors, and charge-density wave
systems. These large characteristic time scales suggest the presence
of many equilibrium or metastable configurations with a distribution
of free-energy barriers separating them.

For free-energy barriers in spin glasses a major complication arises
from the fact that there is no parametrization of the relevant
configurations by a conventional thermodynamic variable. In his
work~\cite{Pa79} on the mean-field theory of spin glasses Parisi
generalized the concept of an order parameter.
In later language~\cite{Young97,Fi91,Me86,Bi86}
this is expressed as follows: A spin-glass realization is defined
by a set of frozen, disordered exchange coupling constants
${\cal J}=\{ J_{ik} \}$ and for each realization the Parisi overlap
parameter is defined by
\begin{equation} \label{q}
q = {1\over N} \sum_{i=1}^N s^1_i\, s^2_i\ ,
\end{equation}
where the sum goes over the total number $N$ of spins of the system
and the spin superscripts label two (real) 
replica of the same realization.
For given ${\cal J}$ the probability density of $q$ is denoted by
$P_{\cal J}(q)$ and its cumulative distribution function is
$x_{\cal J}(q) = \int_{-q}^q dq' P_{\cal J}(q')$. Average
over the disorder defines the functions
$$ P(q) = \left[ P_{\cal J}(q) \right]_{\rm av} =
   {1\over \# J} \sum_{\cal J} P_{\cal J}(q) ~~{\rm and}~~
   x(q) = \left[ x_{\cal J}(q) \right]_{\rm av} =
   {1\over \# J} \sum_{\cal J} x_{\cal J}(q)\ ,$$
where $\# J$ is the number of realizations considered.
In the infinite
volume limit 
below the freezing temperature
an increasing {\it continuous} part of $x(q)$
characterizes mean-field behavior of spin glasses, whereas in
ferromagnets as well as in the droplet picture~\cite{FiHu88} of
spin glasses $x(q)$ is a step function.

Analytical calculations in mean-field theory show that violations
of the fluctuation-dissipation theorem in non-equilibrium dynamics
determine the static function $x(q)$ and vice versa~\cite{CuKu93},
see~\cite{Bo97} for a review.
Numerical
calculations in $3d$ and $4d$ Ising spin glasses~\cite{FaRi95,MaPa98}
support that this relationship holds also in finite dimensions. Of
course, the entire $P_{\cal J}(q)$ set contains more information than
its mean $P(q)$ (equivalently $x(q)$). In this paper we study the
distribution of the minima in $q$ of the $P_{\cal J}(q)$ probability
densities. For given ${\cal J}$ the non-trivial ({\it i.e.} away from
$q=\pm 1$) minima are related to free-energy barriers of the disordered
system ${\cal J}$. The other way round, it is presumably model
dependent (and worthwhile to investigate) to what extent free-energy
barriers of the system ${\cal J}$ are reflected in the minima of
the $P_{\cal J}(q)$ probability density.

Conventional, canonical Monte Carlo (MC) simulations do not allow for an 
efficient
investigation of the $P_{\cal J}(q)$ minima, because the likelihood
to generate corresponding configurations in the Gibbs canonical
ensemble is small. This problem is overcome by the multi-overlap
MC algorithm~\cite{BeJa98} which samples with an uniform distribution
in $q$. It belongs to the class of  multicanonical and related
algorithms~\cite{muca,muma}, which allow to focus on rare
configurations of the Gibbs ensemble. For instance, at first-order 
phase transitions in $3d$, configurations with interfaces 
are suppressed according to $\exp (-\sigma\, A_{\min})$, where 
$\sigma$ is the interface tension and $A_{\min}$ is the minimal
area of the interface. For
temperature driven transitions configurations with interfaces are
found for $E$ in the energy range $E_1 < E < E_2$
where $E_2=E_1+\triangle E$ and $\triangle E$ is the latent 
heat of the transition. To generate such configurations 
with a good statistics it is sufficient to  sample with a weight
factor $w(E) \sim 1/n(E)$, where $n(E)$ is the spectral density. 
Similarly, interfaces for magnetic field driven first-order phase
transitions can be generated by sampling with an appropriate weight
function $w(M)$ of the magnetization $M$ of the sample.

Once $P_{\cal J}(q)$ is determined, we define the associated free-energy 
barrier $F_B^q$ through the autocorrelation time of a $1d$ 
Markov process which has the canonical $P_{\cal J}(q)$ distribution as
equilibrium state. The barrier autocorrelation time $\tau^q_B$ is then
defined through the second largest eigenvalue of the transition matrix
of this Markov process and the free-energy barrier is $\ln (\tau^q_B)$.

In previous literature \cite{MaYo82,Ne88,RoMo89,VeVi89,unpublished,
JBB98} investigations of spin-glass barriers relied on various 
numerical and analytical methods, which are distinct from ours. The
results of \cite{MaYo82,Ne88,RoMo89,VeVi89,unpublished} may be 
summarized as support of a scaling law $F_B^{\rm can} \sim N^{1/3}$ for
canonical free-energy barriers in the mean-field limit below the
freezing temperature.

In the next section we describe our methods and give an overview of
our MC statistics. Section~\ref{sec_Barrier} presents and interprets
our numerical results for free-energy barrier in $q$. Conclusions and
an outlook are given in the final section~\ref{sec_conclude}.

\section{Overview of Methods and Data}

The energy of the Edwards-Anderson Ising (EAI)~\cite{EA75} spin-glass 
model is given by
\begin{equation} \label{energy}
E = - \sum_{\langle ik \rangle} J_{ik}\, s_i s_k\ ,
\end{equation}
where the sum is over nearest-neighbor pairs of a (hyper)
cubic lattice. The spins $s_i$ as well as the coupling constants 
$\displaystyle J_{ik}$ take on the values $\pm 1$, with equal 
probabilities, {\it i.e.} the sum $N^{-1} \sum_{\langle ik \rangle} J_{ik}$ is of 
order $1/\sqrt N$.

In our calculations
we combine the two copies (replica) of the same realization and simulate
with a weight function
\begin{equation} \label{weight}
 w(q) = \exp \left[- \beta (E^1 + E^2) + S(q) \right]\ .
\end{equation}
Here 
$\beta = J_0/k_B T$ is the inverse temperature in natural units,
$E^1$ and $E^2$ are the energies of the respective replicas and 
$S(q)$ has the meaning of the microcanonical entropy of the Parisi order 
parameter (\ref{q}).
The multi-overlap algorithm weights spin configurations with an overlap 
parameter $q$
in such a way that a broad histogram in $q$, eventually covering 
the entire accessible range $-1\le q\le 1$, is obtained. This allows then for 
accurate calculations of the empirical probability density 
$P_{\cal J}(q)$ of the Parisi order parameter for realization 
${\cal J}$. Although an explicit order parameter does not
exist, our simulation method~\cite{BeJa98} is in this way similar
to the multimagnetical~\cite{muma}, which for ferromagnetic systems
is a very efficient way to sample configurations with interfaces. 

Our EAI simulations are performed on $N=L^d$ $(d=3,4)$ lattices at 
$\beta =1\ (3d)$ and $\beta=0.6\ (4d)$. Both values correspond to
temperatures $T=1/\beta$ below the freezing temperature of the
respective model ($\beta_c = 0.90 \pm 0.03 \ (3d)$
\cite{KaYo96}, $\beta_c = 0.485 \pm 0.005 \ (4d)$ \cite{BaDo93}). 
Table~\ref{Stat_Tab} summarizes the statistics we have assembled as
well as the performance of our code. MC updates are given in
units of {\it sweeps}. Our $J_{ik}$ realizations were drawn 
using the pseudo random number generators RANMAR~\cite{Ma90} and 
RANLUX~\cite{Lu94} (luxury level 4). In the simulations themselves 
we always employed the RANMAR generator due to CPU time considerations.

For each realization ${\cal J}$ the simulation consisted of three
steps:
\begin{enumerate}
\item Construction of the weight function (\ref{weight}). Here we
employed an improved variant of the accumulative stochastic 
iteration scheme discussed in Ref.~\cite{Be96}, algorithmic details 
will be published elsewhere~\cite{BBJ2000}. The iteration was 
stopped after at least 4~tunneling events 
\begin{equation} \label{tunnel}
 (q=0) \to (q=\pm 1) ~~{\rm and\ back}
\end{equation}
occurred. Our precise 
request was in $3d$ 10~tunneling events for $L=4$, 6, and 8, and 20 
events for $L=12$, but for a few cases with only 4 events requested. 
In $4d$ it was 10 for $L=4$, 20 for $L=6$ and 20 to 30 for $L=8$.
In few cases, the system was tunneling so rarely between $q=\pm 1$
that we decided to abort the run and restart with a different random 
number seed, which in most cases led (eventually after multiple
tries) to improved tunneling performance.
After the weight function is constructed and kept fixed, the average
number of sweeps it takes to create a tunneling event (\ref{tunnel}) 
defines the autocorrelation time of the multi-overlap algorithm
which in the following is denoted by
\begin{equation} \label{tau_muqu}
\tau^{\rm muq}\ .
\end{equation}
Of course, $\tau^{\rm muq}$ depends on the realization ${\cal J}$ at
hand, and on the parameters used in phase 1: random number seed, number
of tunneling events requested, etc..
\item Equilibration run. This run of $n\times 65536$ sweeps was done 
to equilibrate the system for given fixed weight factors 
($n=1,4,16,32$ for $3d$ $L=4,6,8,12$ and $n=2,8,16$ for $4d$ 
$L=4,6,8$, respectively).
\item Production run. Each production run of data taking was concluded 
after at least 20 tunneling events as defined in equation~(\ref{tunnel})
were recorded.  To allow for standard reweighting in temperature we 
stored besides histograms of the Parisi overlap parameter also a time 
series of  measurements for the order parameter, energies and
magnetizations of the two replica. 
The number of sweeps between two successive points in a time
series is adjusted in such a way that each time series is made
of 65536 measurements. This is done by 
an adaptive data compression routine~\cite{BBJ2000}. Together
with the condition on the minimal number of tunneling events this
ensures that the number of sweeps between two successive points in a 
time series is approximately  
proportional to $\tau^{\rm muq}$.
Some re-weighting results were reported in
references~\cite{BeJa98} and~\cite{JBB98}, publication of others 
is intended~\cite{BBJ2000}. 

\end{enumerate}

With each realization ${\cal J}$ we associate the free-energy 
barrier $F_B^q$ of the $1d$ Metropolis-Markov chain~\cite{Me53} which
has the canonical $P_{\cal J}(q)$ probability density as its 
equilibrium distribution. The transition probabilities $T_{i,j}$ 
are given by
\begin{equation} \label{T}
T=\left[\begin{array}{cccc}1-w_{2,1}&w_{1,2}&0 & \ldots\cr
	    w_{2,1}&1-w_{1,2}-w_{3,2}&w_{2,3}&\ldots\cr
	    0&w_{3,2}&1-w_{2,3}-w_{4,3} &\ldots\cr
            0 & 0 & w_{4,3} & \ldots\cr
            \vdots&\vdots&\vdots&\ddots\cr\end{array}\right]\ ,
\end{equation}
where $w_{i,j}$ ($ i \neq j)$ is a   probability \`a la Metropolis 
to jump from state $q=q_j$ to $q=q_i$ ($q_i=i/N$, 
$i\in [-N , -N+2 , \dots +N]$),
\begin{equation} 
 w_{i,j}={1\over 2} 
 \min\Bigl(1,{P_{\cal J}(q_i)\over P_{\cal J}(q_j)}\Bigr)\ .
\end{equation}  
$T$  fulfills the detailed balance condition (with $P_{\cal J}$) and
as a consequence it has only real eigenvalues. The largest eigenvalue 
(equal to one) is non-degenerate, and the second largest eigenvalue 
$\lambda_1$ determines the autocorrelation time of the chain,
\begin{equation} \label{tauq}
\tau_B^q = {1\over{N (1-\lambda_1)}}\ ,
\end{equation} 
and we define the associated free-energy barrier for realization
${\cal J}$ as
\begin{equation} \label{F_B^q}
 F_B^q = \ln (\tau_B^q)\ .
\end{equation}
For the simple double-peak situation of first-order phase transitions 
the autocorrelation time $\tau^q_B$ is proportional to the ratio
$P_{\cal J}^{\max}/P_{\cal J}^{\min}$ where
$$ P_{\cal J}^{\max} = P_{\cal J} (q_{\max} ) = \max_q 
  \left[ P_{\cal J}(q) \right] ~~{\rm and}~~ P_{\cal J}^{\min} 
  = \min_{0<q<|q_{\max}|} \left[ P_{\cal J}(q) \right]\ , $$
{\it i.e.} $F_B^q = \ln [ P^{\max}_{\cal J} / P^{\min}_{\cal J}] + const.$
This leads in $3d$ to $\tau^q_B \sim e^{\sigma A_{\min}}$ and 
$F_B^q \sim \sigma A_{\min} + const.$, where $A_{\min}$ is the minimal
area between the coexisting phase regions and $\sigma$ is the interfacial
tension.
Equation (\ref{F_B^q}) is the appropriate generalization to a situation
involving multiple minima and maxima. The autocorrelation time
$\tau^q_B$ has
to be regarded as a lower limit to the canonical autocorrelation time
$\tau^{\rm can}$ for the Markov process where the spin variables are the
dynamical degrees of freedom. The definition~(\ref{tauq}) takes only
barriers in $q$ into account but not other barriers which may well
exist in the multi-dimensional configuration space. 
\begin{figure}[tb]
\vspace{1pc}
 \centerline{\hbox{ \psfig{figure=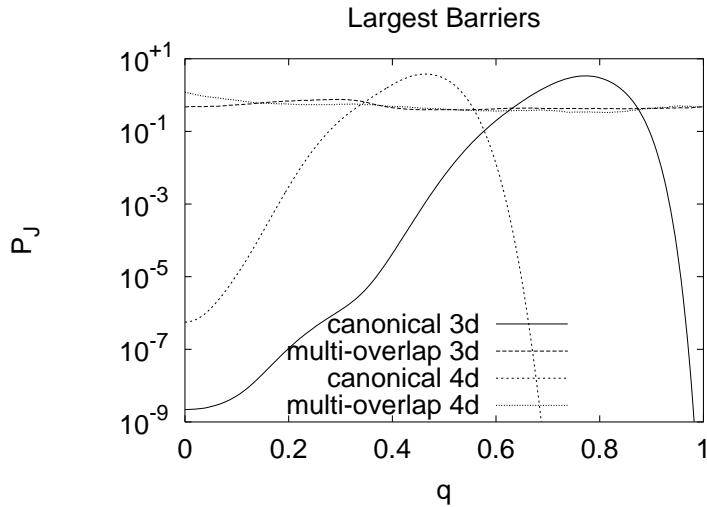,width=10cm} }}
 \caption{Canonical $P_{\cal J}(q)$ and (flat) multi-overlap
 $P^{\rm muq}_{\cal J}(q)$ probability densities for our realization
 with the largest free-energy barrier in $3d$ ($L=8$) and $4d$ ($L=8$).}
 \label{fig_PJq}
\end{figure}

The matrix $T$ in (\ref{T}) is tri-diagonal and sign symmetric. This 
special form allows for easy calculation of all its 
eigenvalues~\cite{Eislib}.
The realizations with the largest thus obtained free-energy barriers
in $3d$ and $4d$ are depicted in figure~\ref{fig_PJq}. 
Both do not show a complicated landscape, but a plain two-peak
structure. Besides the
canonical $P_{\cal J}(q)$ probability densities the essentially flat 
probability densities $P_{\cal J}^{\rm muq} (q)$ of the multi-overlap 
simulation are also indicated in the figure.
Both $P_{\cal J}(q)$ probability densities take their minimum at
$q=0$ and we have $P_{\cal J}^{\max}/P_{\cal J}^{\min} > 10^8$ in $3d$ ($L=8$)
and $P_{\cal J}^{\max}/P_{\cal J}^{\min} > 10^6$ in $4d$ ($L=8$); compare
also table 2. 
The improvement factors in computer time are directly proportional
and close to these numbers which reflect the enhancements in visits
of $P_{\cal J}(0)$ as compared to canonical simulations.
Multiplying the improvement factors with the average CPU times needed
by the multi-overlap algorithm for a single realization on lattices of
this size (see table~\ref{Stat_Tab}), it becomes clear that exploration
of such barriers by means of a canonical MC simulation is simply
impossible.

\begin{figure}[tb]
\vspace{1pc}
 \centerline{\hbox{ \psfig{figure=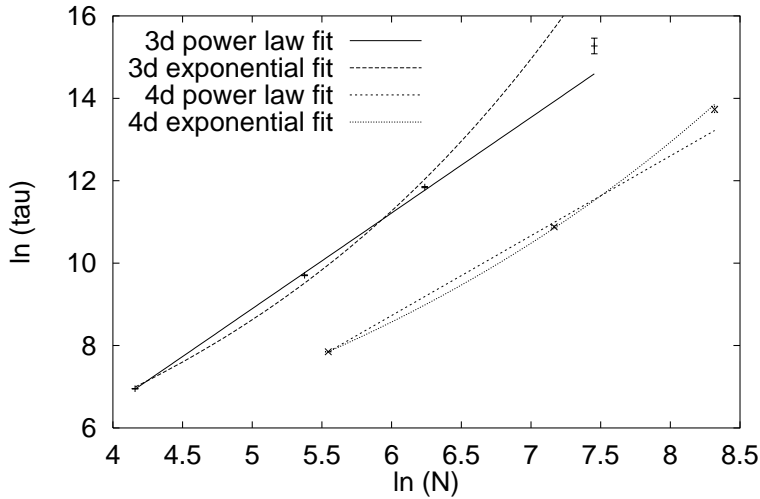,width=10cm} }}
 \caption{Power law and exponential fits for the mean multi-overlap 
 tunneling time $[\tau^{\rm muq}]_{\rm av}$ in $3d$ and $4d$.}
 \label{fig_tun_fits}
\end{figure}

We conclude this section with remarks about the performance of
the algorithm and implications on the physics of the system. 
The multi-overlap algorithm flattens
the free-energy barriers $F^q_B$. 
If they were the only cause for the slowing down of
the canonical dynamics, the multi-overlap autocorrelation time should
be dominated by a random walk behavior between $q=-1$ and $q=+1$ and
scale proportional to $N$ (in units of sweeps).
Fitting the estimates of the mean autocorrelation time
$[\tau^{\rm muq}]_{\rm av}$, where the average is with respect to the
realizations ${\cal J}$, to the power-law form
$\ln ([\tau^{\rm muq}]_{\rm av}) = a+z\,\ln (N)$
gives $z=2.32\pm 0.07$ in $3d$ and $z=1.94\pm 0.02$ in $4d$.
The fits are depicted in figure~\ref{fig_tun_fits}. Their quality is 
bad, nevertheless they show that the slowing down is quite off from
the theoretical optimum $z=1$. Exponential fits 
$\ln ([\tau^{\rm muq}]_{\rm av}) = c_0+c_1 N$
are also depicted in the figure. Whereas in $3d$ the exponential fit is 
far worse than the power-law fit, it is the other way round in $4d$.
Hence, the smaller $z$-value in $4d$ should not be taken seriously.

The physically important conclusion is: the observed large 
autocorrelation times demonstrate that, in the model considered,
canonical overlap barriers are not an exclusive cause for the
slowing down of spin-glass dynamics below the freezing temperature.
Therefore, $\tau_B^q$ has to be
a lower bound of the full canonical autocorrelation time $\tau^{\rm can}$:
\begin{equation} \label{tauc}
 \tau^q_B < \tau^{\rm can}\ . 
\end{equation}
One should understand $q$ as one relevant direction in a complex,
multidimensional configuration space. By depicting free-energy
barriers as function of $q$ one projects on this direction and
averages results over all other directions. 

\section{Barrier Results} \label{sec_Barrier}

We analyze our free-energy barrier densities relying on a variant of the 
cumulative distribution function $F$. For a set of sorted data
\begin{equation} \label{sort}
 x_1 < x_2 < \dots < x_n
\end{equation}
the (empirical) cumulative distribution function $F(x)$, see for 
instance~\cite{NumRec}, is defined by 
\begin{equation} \label{F}
 {i\over n} - {1\over 2n} \le F(x)\le {i\over n} + {1\over 2n}
 ~~{\rm for}~~ x_i \le x \le x_{i+1}\ , 
\end{equation}      
where we use a straight-line interpolation in-between.
Next we define a $Q$-tile~\cite{footnote1} distribution function
as introduced in~\cite{Be_book}
\begin{equation} \label{Fq}
 F_Q (x) = \cases{ F(x)\ {\rm for}\ F(x)\le 0.5\ ;\cr
               1 - F(x)\ {\rm for}\ F(x)\ge 0.5\ .\cr}
\end{equation}
This function peaks at the median $x_{\rm med}$ of the data and 
takes there the value $F_Q=0.5$. For self-averaging data $x$ 
the function $F_Q$ collapses in the infinite volume to
$$ F_Q (x) = \cases{ 0.5\ {\rm for}\ x=\overline{x}\ ;\cr
                     0\ {\rm otherwise}\ .\cr} $$
Here $\overline{x}$ is the mean value. For non-averaging 
quantities the width of $F_Q$ stays finite. The concept carries 
over to observables which diverge in the infinite volume limit,
when on each lattice size results are expressed in units of the
respective median value, {\it i.e.} instead of an observable
$X$ the ratio $x=X/X_{\rm med}$ is used.

\subsection{Lack of self-averaging}

\begin{figure}[tb]
\vspace{1pc}
 \centerline{\hbox{ \psfig{figure=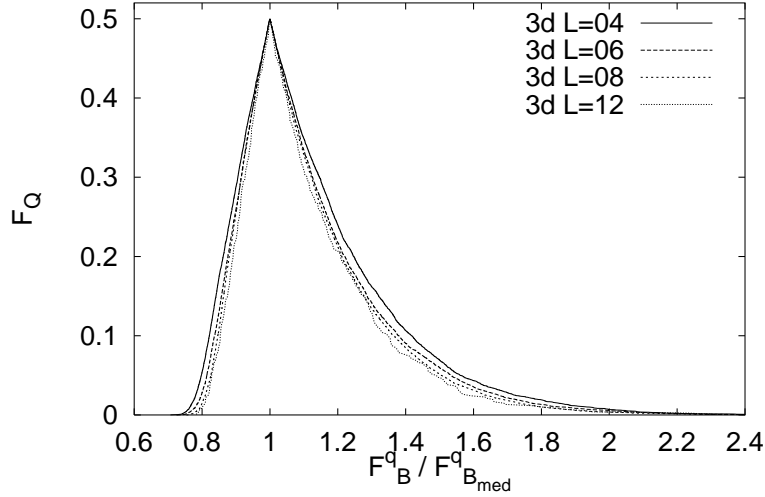,width=10cm} }}
\caption{Distribution function $F_Q$~(\ref{Fq}) for the $3d$ overlap
barriers~(\ref{F_B^q}) in units of their median value.
} \label{fig_FqB3d}
\end{figure}
\begin{figure}[tb]
\vspace{1pc}
 \centerline{\hbox{ \psfig{figure=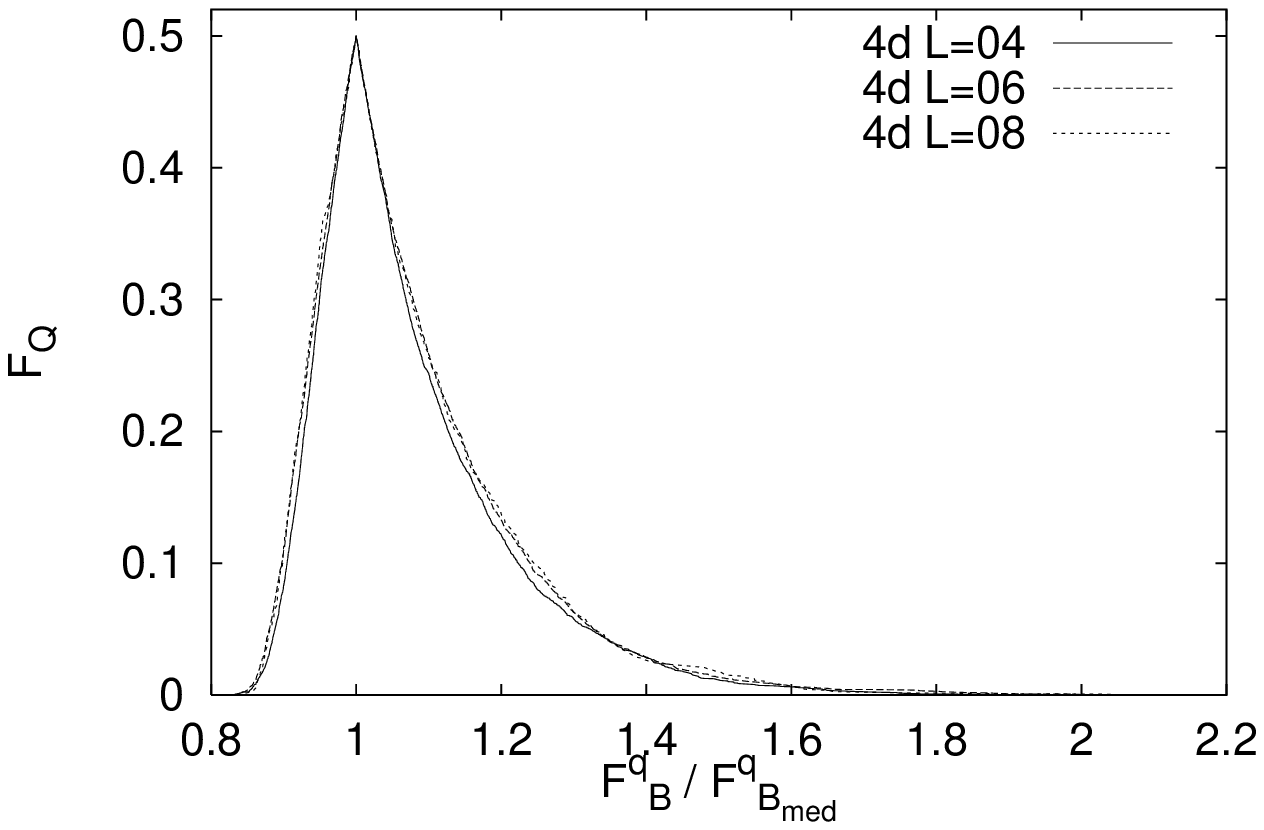,width=10cm} }}
\caption{Distribution function $F_Q$~(\ref{Fq}) for the $4d$ overlap
barriers~(\ref{F_B^q}) in units of their median value.
} \label{fig_FqB4d}
\end{figure}
\begin{figure}[tb]
\vspace{1pc}
 \centerline{\hbox{ \psfig{figure=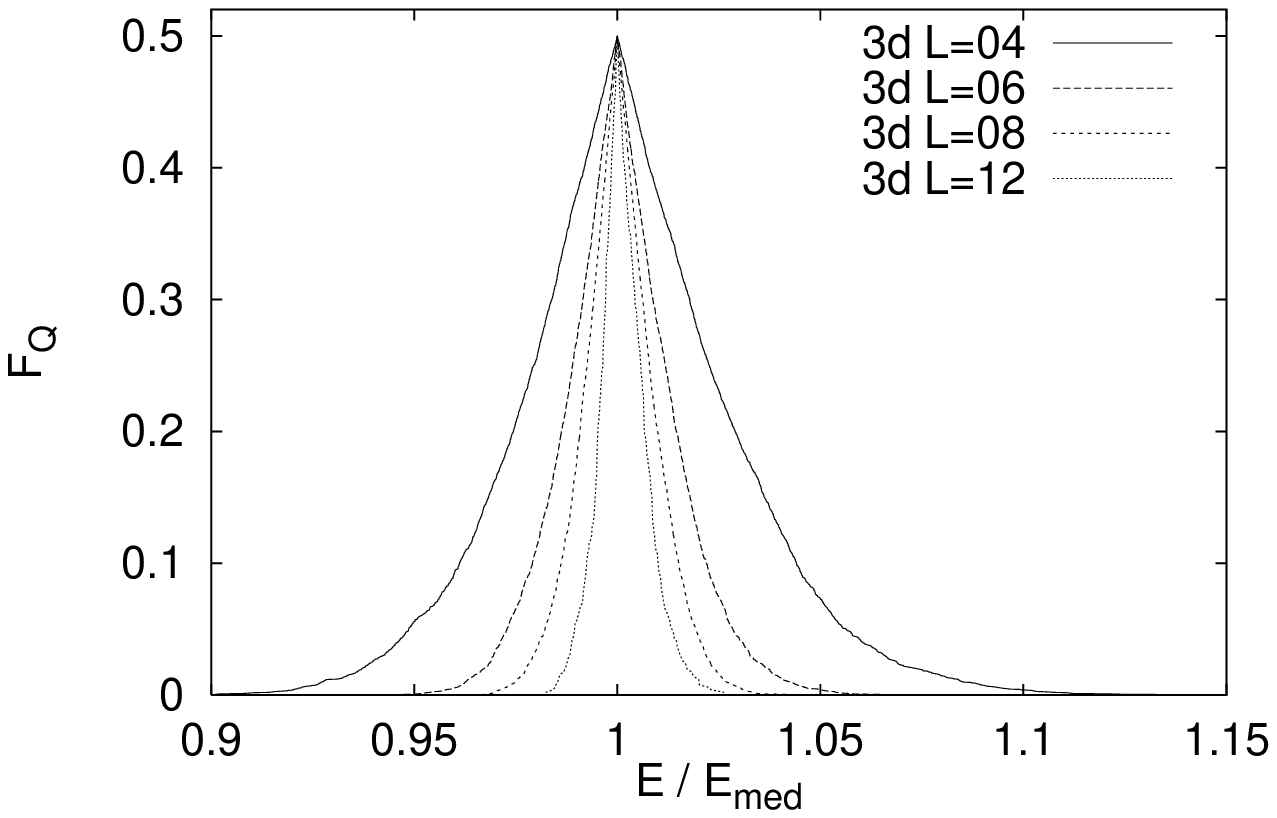,width=10cm} }}
\caption{Distribution function $F_Q$~(\ref{Fq}) for the $3d$
energies~(\ref{energy}) in units of their median value.
} \label{fig_FqE3d}
\end{figure}
\begin{figure}[tb]
\vspace{1pc}
 \centerline{\hbox{ \psfig{figure=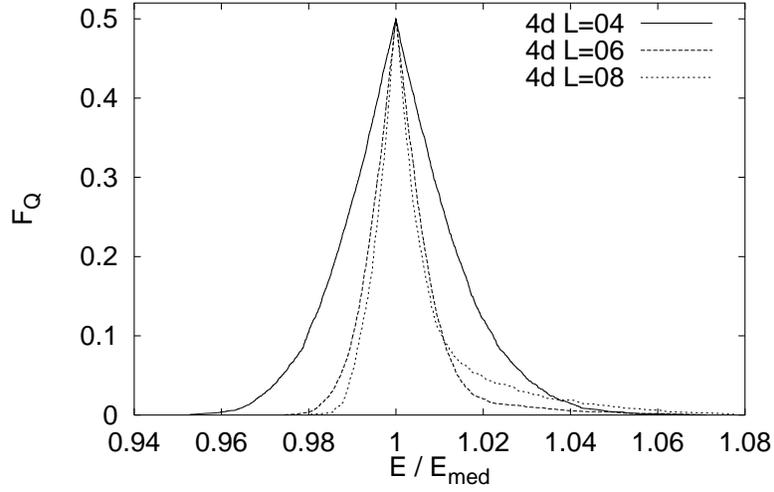,width=10cm} }}
\caption{Distribution function $F_Q$~(\ref{Fq}) for the $4d$
energies~(\ref{energy}) in units of their median value.
} \label{fig_FqE4d}
\end{figure}

For the free-energy barriers (\ref{F_B^q}) we have depicted
our thus obtained $F_Q(F_B^q/F^q_{B,\rm med})$ functions in
figures~\ref{fig_FqB3d} ($3d$) and~\ref{fig_FqB4d} ($4d$).
For each lattice the measured $F_B^q$ values were first
sorted as function of ${\cal J}$ such that
$$  F^q_{B,1} < F^q_{B,2} < \dots < F^q_{B,n}\ ,$$
where $n$ is the number of realizations $\# {\cal J}$ given
in table~\ref{Stat_Tab}. Subsequently $F^q_{B,\rm med}$ was
calculated as
$$ F^q_{B,\rm med} = {1\over 2} \left( F^q_{B,n/2} +
   F^q_{B,1+n/2} \right) ~~~(n\ {\rm is\ even\ in\ our\ cases})\ ,$$
and $F_Q$ computed for $x=F^q_B/F^q_{B,\rm med}$.

Both figures support that $F^q_B$ is a non-self-averaging quantity.
This is stronger in $4d$ than in $3d$, because the inner lines
belong in $3d$ to the larger lattices, whereas in $4d$ it
is the other way round. However, in both cases there are marginal
finite-size effects, whereas finite-size dependence of self-averaging
is expected
to be rather strong. This becomes obvious when comparing with an 
observable which is supposed to be self-averaging. Namely, 
figures~\ref{fig_FqE3d} ($3d$) and~\ref{fig_FqE4d} ($4d$) 
depict the same analysis for the internal energy~(\ref{energy}).
In $3d$ self-averaging of this quantity is obvious, whereas 
in $4d$ there is an irregularity when going from $L=6$ to 
$L=8$. As our simulation temperature in $4d$ is quite low, we 
think that this behavior is related to groundstate 
irregularities on small lattices (only the corresponding half 
of the distribution is affected).
For both $3d$ and $4d$ the $q$-tile distribution function of the energy
is strongly peaked around  $E/E_{\rm med}=1$, whereas
the overlap barrier distributions are much broader.

It is generally believed that, in contrast to the equilibrium 
autocorrelation times considered here, non-equilibrium
autocorrelations are self-averaging~\cite{Bo97}.
No sample-to-sample deviations have been reported for real 
experiments~\cite{Gr87} and self-averaging is also used for 
measurements of non-equilibrium properties in MC
simulations~\cite{FaRi95,MaPa98}.

\begin{figure}[tb]
\vspace{1pc}
 \centerline{\hbox{ \psfig{figure=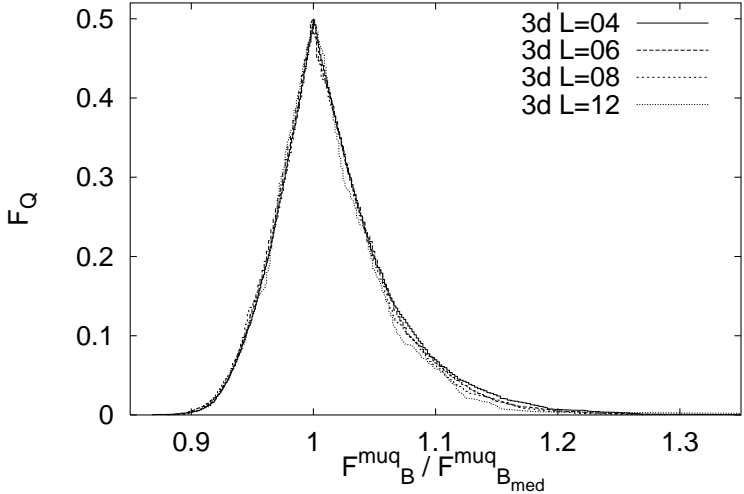,width=10cm} }}
\caption{Distribution function $F_Q$~(\ref{Fq}) for the $3d$
barriers~(\ref{F_B^muq}) of the multi-overlap algorithm in
units of their median value.
} \label{fig_FqT3d}
\end{figure}
\begin{figure}[tb]
\vspace{1pc}
 \centerline{\hbox{ \psfig{figure=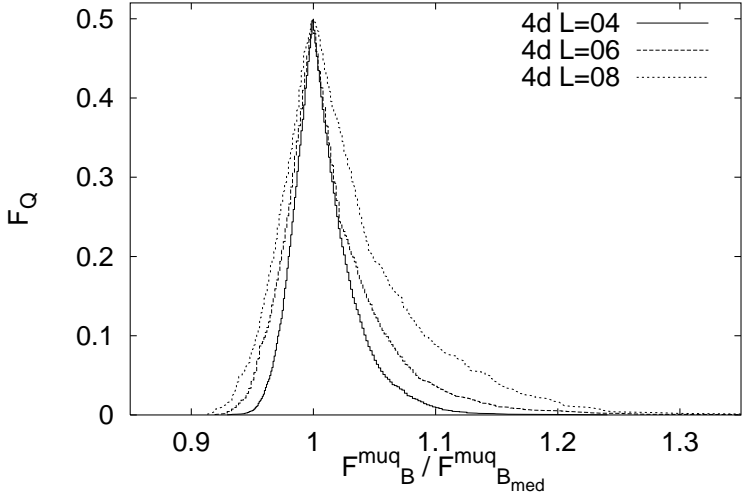,width=10cm} }}
\caption{Distribution function $F_Q$~(\ref{Fq}) for the $4d$
barriers~(\ref{F_B^muq}) of the multi-overlap algorithm in
units of their median value.
} \label{fig_FqT4d}
\end{figure}

The multi-overlap algorithm eliminates the free-energy
barriers which are visible in the $P_{\cal J}(q)$ probability
densities. Let us therefore focus on the autocorrelations times 
of this algorithm and its barriers defined by
\begin{equation} \label{F_B^muq}
F_B^{\rm muq} = \ln (\tau^{\rm muq})\ .
\end{equation}
We perform the analysis along our previous lines and show in
figures~\ref{fig_FqT3d} ($3d$) and~\ref{fig_FqT4d} ($4d$) the thus
obtained $F_Q(F_B^{\rm muq}/F^{\rm muq}_{B,\rm med})$ functions. Lack of
self-averaging is even more obvious than for $F^q_B$. In
figure~\ref{fig_FqT3d} ($3d$) there are (within the statistical
accuracy) no finite-size effects visible and
figure~\ref{fig_FqT4d} ($4d$) exhibits a strong
anti-self-averaging trend: Results for the larger lattices move
to the outside instead to the inside. 

\subsection{Finite-size scaling behavior}

In this final part of section~\ref{sec_Barrier} we discuss how data 
(experimental or MC)
for non-self-averaging observables may be analyzed such that
comparisons of results from different groups become possible. 
One has to investigate many samples and should report
the finite-size scaling behavior for fixed values of the cumulative
distribution function $F$~(\ref{F}). In particular this includes 
$F=1/2$ which defines the median value.
We exemplify this for the overlap autocorrelation time
$\tau_B^q$~(\ref{tauq}), but the method applies for
non-self-averaging observables in general.

From figures~\ref{fig_FqB3d} and~\ref{fig_FqB4d} it is obvious that
the autocorrelation times $\tau_B^q$ will have long tails towards
large values. This implies
that the mean value over all samples is a rather erratic
quantity which is dominated by a few rare realizations.
Table~\ref{Barrier_Tab} collects the mean, median ($F=1/2$) and
maximum ($F=1-1/(2n)$) values for $\tau_B^q$.  The numbers in
parenthesis indicate error bars in the last digits of the
quantity given before. The results show that contributions of
the maximum values dominate to a large extent the mean values
(just divide the maximum values by the number of realizations
and compare the results with the mean values).
The maximum values rely on realizations ${\cal J}$
of likelihood $1/n$, what explains their very large errors.
In contrast to the mean and the maximum, results for fixed $F$ remain
well defined as long as $F$ stays away from its extreme values
$1/n$ and $1-1/n$. In particular, note that samples with
relaxation times too long to be measured can still contribute to
determine the correct $F$ values for smaller relaxation times.

\begin{figure}[tb]
\vspace{1pc}
 \centerline{\hbox{ \psfig{figure=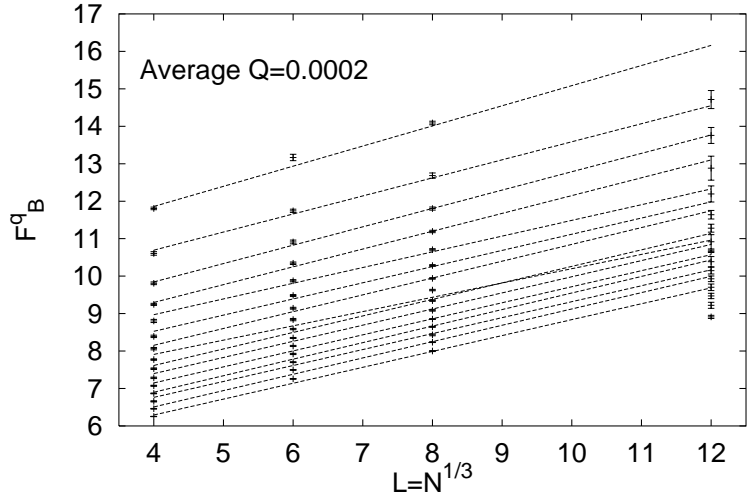,width=10cm} }}
\caption{Fits (\ref{F_Bfits1}) of the $3d$ free-energy barriers $F_B^q$
versus $N^{1/3}$ corresponding to the exponential finite-size scaling behavior
(\ref{tau_exp}) of $\tau_B^q$. From down to up the lines are at
$16\,F = 1,3,5,7,9,11,12,13,14$, and $15$.
} \label{fig_3d_expfits}
\end{figure}
\begin{figure}[tb]
\vspace{1pc}
 \centerline{\hbox{ \psfig{figure=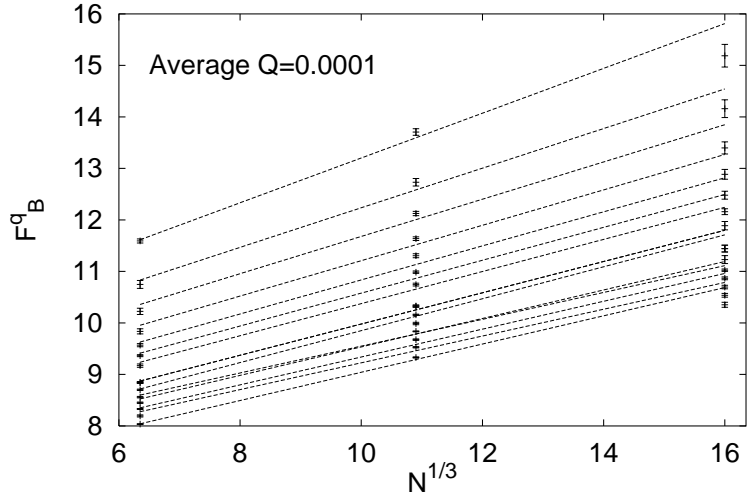,width=10cm} }}
\caption{Fits (\ref{F_Bfits1}) of the $4d$ free-energy barriers $F_B^q$
versus $N^{1/3}$ corresponding to the exponential finite-size scaling behavior
(\ref{tau_exp}) of $\tau_B^q$. From down to up the lines are at
$16\,F = 1,3,5,7,9,11,12,13,14$, and $15$.
} \label{fig_4d_expfits}
\end{figure}

\begin{figure}[tb]
\vspace{1pc}
 \centerline{\hbox{ \psfig{figure=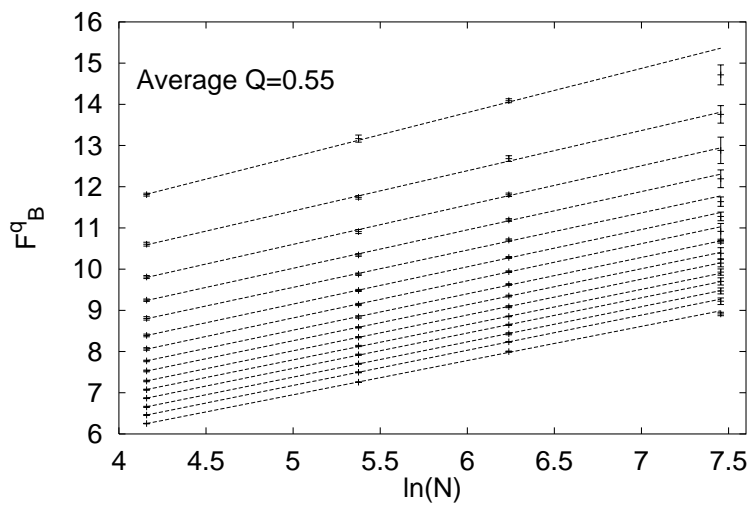,width=10cm} }}
\caption{Fits (\ref{F_Bfits2}) of the $3d$ free-energy barriers $F_B^q$
versus $\ln(N)$ corresponding to the power-law finite-size scaling behavior
(\ref{tau_power}) of $\tau_B^q$. From down to up the lines are at
$16\,F = 1,3,5,7,9,11,12,13,14$, and $15$.
} \label{fig_3d_powfits}
\end{figure}

\begin{figure}[tb]
\vspace{1pc}
 \centerline{\hbox{ \psfig{figure=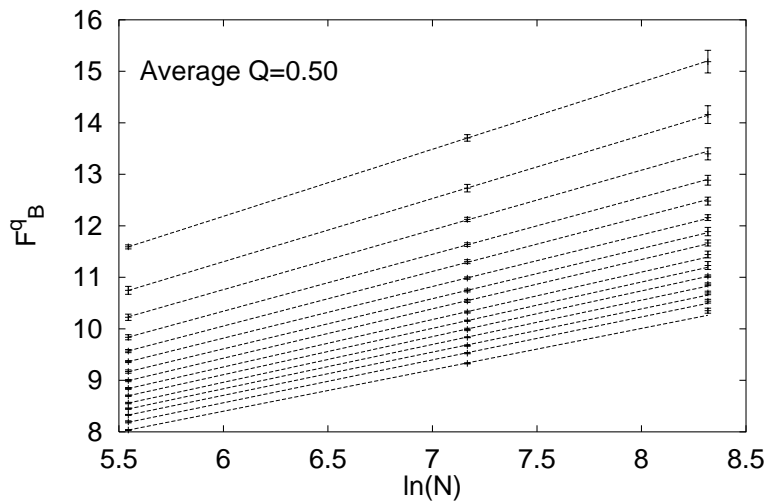,width=10cm} }}
\caption{Fits (\ref{F_Bfits2}) of the $4d$ free-energy barriers $F_B^q$
versus $\ln(N)$ corresponding to the power-law finite-size scaling behavior
(\ref{tau_power}) of $\tau_B^q$. From down to up the lines are at
$16\,F = 1,3,5,7,9,11,12,13,14$, and $15$.
} \label{fig_4d_powfits}
\end{figure}

In the following we focus on our results for the free-energy
barriers $F_B^q$ at $F=i/16$ with $i=1,\dots ,15$. For each
$F$ value we performed fits to the form 
\begin{equation} \label{F_Bfits1}
   F_B^q = a_1 + a_2\, N^{1/3}   
\end{equation}
which corresponds to the exponential finite-size scaling behavior
\begin{equation} \label{tau_exp}
 \tau_B^q = e^{a_1}\, e^{a_2\, N^{1/3}}
\end{equation}
suggested by investigations of autocorrelation times and barriers
in the mean-field limit~\cite{MaYo82,Ne88,RoMo89,VeVi89,unpublished}. 
These fits are depicted in figures~\ref{fig_3d_expfits} 
and~\ref{fig_4d_expfits}. Examples of the fit parameters $a_1$
and $a_2$ are collected in table~\ref{Fits1_Tab}; for all fits 
given there, the goodness-of-fit parameter $Q$~\cite{NumRec} is smaller
than 0.003. The average $Q$ over all 15 fits is given in the figures.
For consistent fits the expectation for the $Q$-average
is $1/2$ and the quality of our $3d$ and $4d$ exponential fits is
unacceptable. We therefore try a power-law fit
\begin{equation} \label{tau_power}
 \tau_B^q = c\, N^{\alpha}\ ,
\end{equation}
which corresponds to a fit of the form
\begin{equation} \label{F_Bfits2}
   F_B^q = \ln (c) + \alpha\, \ln (N)\ .   
\end{equation}
These fits are depicted in figures~\ref{fig_3d_powfits} 
and~\ref{fig_4d_powfits}.  In $3d$ as well as in $4d$
the average $Q$-value is now almost perfect. Examples of the power-law
fit parameters and $Q$-values are given in table~\ref{Fits2_Tab}. They
indicate that the distribution of $Q$-values is
less perfect than their mean.  Such uncertainties are
an intrinsic limitation of MC simulations and become particularly severe
when one is, as in our investigation, limited to rather small-sized
systems. Having these limitations in mind, the over-all quality of the
power-law fits is remarkably good. Our data favor them strongly over
the exponential behavior.


As function of $F$ the exponent $\alpha = \alpha (F)$ varies smoothly
and covers in $4d$ a range from 0.8 at $F=1/15$ to 1.3 at
$F=15/16$. In $3d$ the range is somewhat smaller, see
table~\ref{Fits2_Tab}. Fits for $F> 15/16$
become erratic and it makes little sense to report them. A similar
analysis for the autocorrelation times of the multi-overlap algorithm
gives exponents $\alpha (F)$ which are larger,
$$ \alpha^{\rm muq} (F) \approx \alpha^q_B (F) +1\ .$$
This re-iterates and sharpens our previous observation that relevant
barriers exist, which are invisible in the overlap variable $q$.

\section{Summary and Conclusions} \label{sec_conclude}

We have investigated free-energy barriers in the Parisi order
parameter (\ref{q}). The results are sample dependent and
non-self-averaging on the (admittedly rather small)
simulated systems. The power-law behavior (\ref{tau_power}) 
of the Markov autocorrelation
times $\tau^q_B$ as defined in eq.~(\ref{tauq})
is favored over the exponential behavior (\ref{tau_exp}).
To the extent that this behavior extrapolates to the infinite
volume limit and that our methods relate to those of
Ref.~\cite{MaYo82,Ne88,RoMo89,VeVi89,unpublished}, it means
that both $3d$ and $4d$ are quite far away from the $d\to\infty$
mean-field theory limit. As relevant barriers are still found
in the autocorrelations of the multi-overlap algorithm, such
a relation is far from clear.

\section*{Acknowledgements}
B.B. likes to thank Wolfgang Grill for his hospitality during an extended 
stay at Leipzig University in the framework of the {\em International Physics 
Studies Program\/}.
Most numerical simulations were performed on the T3E computer of CEA 
in  Grenoble under grant p526, additional calculations were done on 
T3E computers of ZIB in Berlin under grant bvpl01, and of NIC in 
J\"ulich under grant hmz091, and on Alpha workstations at FSU.  We 
thank all institutions for their generous support.

\bibliographystyle{unsrt}




\singlespace    

\addtolength{\textheight}{0.6cm}
\clearpage
\vspace*{-0.6cm}
\noindent {\Large \bf Tables}



\begin{table}[h]
\caption{\emph{Statistics: Number of realizations \#$\cal J$, average
number of mega-sweeps per realization $n_{\rm sw}$ and average single
375~MHz processor CPU time per realization in hours (h) or seconds (s)
as benchmarked on the CEA T3E. \protect\label{Stat_Tab}}}
\medskip
\smallskip
\centering
\begin{tabular}{||r|c|c|c|c|c|c||}                        \hline
   & 
\multicolumn{3}{|c|}{$3d$} &
\multicolumn{3}{|c||}{$4d$}    \\ \hline
\multicolumn{1}{||c|}{$L$} & \#$\cal J$ & $n_{\rm sw}$ & CPU
   & \#$\cal J$ & $n_{\rm sw}$ & CPU          \\ \hline
 4 & 8\,192 & ~~0.2 M& ~6.32 s& 4\,096 &\ 0.4 M& 76.6 s \\ \hline
 6 & 8\,192 & ~~1.0 M&~\,113 s& 4\,096 &\ 3.7 M& 1.02 h \\ \hline
 8 & 8\,192 & ~~7.6 M& $\!$~0.54 h& 1\,024 & $\!49.3$ M& $\!\!42.66$ h \\ \hline
12 & ~~640 & $\!\!154.0$ M& $\!$36.97 h&       &       &         \\ \hline
\end{tabular}
\vspace*{0.2cm}

\caption{\emph{Mean, median and maximum values for the Markov
autocorrelations times $\tau_B^q$~(\ref{tauq}).
\protect\label{Barrier_Tab}}}
\smallskip
\centering
\begin{tabular}{||l|r|r|r|r||}                                      \hline
 \multicolumn{1}{||c|}{$L$}         & 
\multicolumn{1}{c|}{$4$} & 
\multicolumn{1}{c|}{$6$} & 
\multicolumn{1}{c|}{$8$} & 
\multicolumn{1}{c||}{$12$}                            \\ \hline
$3d$: Mean   &  61(29) $\times 10^4$ &  10(06) $\times 10^6$ &
                56(45) $\times 10^6$ &  13(10) $\times 10^7$         \\ \hline
$3d$: Median & 237(05) $\times 10^1$ & 690(02) $\times 10^1$ & 
               152(04) $\times 10^2$ & 444(05) $\times 10^2$         \\ \hline
$3d$: Maximum&  22(20) $\times 10^8$ &  32(06) $\times 10^9$ &  
                35(33) $\times 10^{10}\!\!$ &  53(30) $\times 10^9$  \\ \hline
$4d$: Mean   &  94(34) $\times 10^3$ &  23(08) $\times 10^5$ &  
                26(23) $\times 10^6$&  \\ \hline
$4d$: Median & 807(18) $\times 10^1$ & 379(11) $\times 10^2$ & 
               117(07) $\times 10^3$&  \\ \hline
$4d$: Maximum&  13(11) $\times 10^7$ &  21(07) $\times 10^8$ &  
                22(21) $\times 10^9$&  \\ \hline
\end{tabular}
\vspace*{0.2cm}

\caption{\emph{Fit parameters for the free-energy barriers 
fits (\ref{F_Bfits1}); $Q < 0.003$ for all cases. \protect\label{Fits1_Tab}} }
\medskip
\smallskip
\centering
\begin{tabular}{||r|c|c|c|c||}                                 \hline
     &  
\multicolumn{2}{|c|}{$3d$}   &  
\multicolumn{2}{|c||}{$4d$}   \\ \hline
 \multicolumn{1}{||c|}{$F$} &  $a_1$  &  $a_2$    &  $a_1$   &   $a_2$    
  \\ \hline
 1/16 & 4.59(2) & 0.424(03) & 6.30(02) & 0.274(03) \\ \hline 
 4/16 & 5.13(3) & 0.442(04) & 6.77(03) & 0.276(03) \\ \hline 
 8/16 & 6.38(4) & 0.382(05) & 7.13(05) & 0.301(06) \\ \hline 
12/16 & 7.39(6) & 0.476(11) & 7.77(11) & 0.344(10) \\ \hline 
15/16 & 9.71(9) & 0.538(14) & 8.85(12) & 0.435(15) \\ \hline 
 \end{tabular}
\vspace*{0.2cm}

\caption{\emph{Free energy barriers fits (\ref{F_Bfits2}): Fit parameters and
goodness of fit. \protect\label{Fits2_Tab}} }
\medskip
\smallskip
\centering
\begin{tabular}{||r|c|c|c|c|c|c||}                              \hline
     &
\multicolumn{3}{|c|}{$3d$}     &   
\multicolumn{3}{|c||}{$4d$}\\ \hline
 \multicolumn{1}{||c|}{$F$} &$\ln (c)$&  $\alpha$ &$Q$ & $\ln (c)$& $\alpha $ &
 $Q$\\ \hline
 1/16 & 2.80(03) & 0.830(06) & 0.04 & 3.58(05) & 0.804(08) & 0.05 \\ \hline
 4/16 & 3.30(04) & 0.857(08) & 0.77 & 3.68(06) & 0.860(08) & 0.13 \\ \hline
 8/16 & 4.10(06) & 0.883(10) & 0.52 & 3.68(11) & 0.958(18) & 0.71 \\ \hline
12/16 & 5.37(11) & 0.930(20) & 0.51 & 3.71(22) & 1.105(32) & 0.81 \\ \hline
15/16 & 7.35(14) & 1.075(28) & 0.02 & 4.37(26) & 1.302(42) & 0.94 \\ \hline
 \end{tabular}
\end{table}


\clearpage
\end{document}